\documentclass[11pt]{article}
\usepackage{amsmath}
\usepackage{graphicx,psfrag,epsf}
\usepackage{enumerate}
\usepackage{natbib}
\usepackage{url}  
\usepackage{tikz}
\usepackage{graphicx}
\usepackage{listings}
\usepackage{enumitem}
\usepackage[normalem]{ulem}
\usepackage{amsmath,mathtools}

\usepackage{units}
\usepackage{bigints}
\usepackage{pdfpages}
\usepackage{setspace}
\usepackage{subcaption}
\usepackage{todonotes}
\usepackage{authblk}

\newcommand{\blind}{0}

\newcommand{\overbar}[1]{\mkern 1.5mu\overline{\mkern-1.5mu#1\mkern-1.5mu}\mkern 1.5mu}

\begin{document}
	
\if0\blind
{
	\title{\bf Bayesian Latent-Normal Inference for the \\Rank Sum Test, the Signed Rank Test, and Spearman's $\rho_s$}
	\author[1]{Johnny van Doorn\thanks{Correspondence may be addressed to Johnny van Doorn (E-mail: \textit{J.B.vanDoorn@uva.nl}), Department of Psychological Methods, University of Amsterdam, Valckeniersstraat 59, 1018XE Amsterdam, the Netherlands. 
			This work was supported in part by a Vici grant from the Netherlands Organization of Scientific Research (NWO) awarded to EJW (016.Vici.170.083). MM is supported by a Veni grant (451-17-017) from the NWO. Centrum Wiskunde \& Informatica is the national research institute for mathematics and computer science in the Netherlands.}}
	
	\author[1,2]{Alexander Ly }
	\author[1]{Maarten Marsman}
	\author[1]{E.--J. Wagenmakers}
	
	\affil[1]{University of Amsterdam}
	\affil[2]{Centrum Wiskunde \& Informatica}
	\setcounter{Maxaffil}{0}
	\date{}
	\maketitle
	
} \fi

\if1\blind
{
	\bigskip
	\bigskip
	\bigskip
	\begin{center}
		{\LARGE\bf Bayesian Rank-Based Hypothesis Testing for the Rank Sum Test, the Signed Rank Test, and Spearman's $\rho_s$}
	\end{center}
	\medskip
} \fi

\bigskip
\begin{abstract}
	Bayesian inference for rank-order problems is frustrated by the absence of an explicit likelihood function. This hurdle can be overcome by assuming a latent normal representation that is consistent with the ordinal information in the data: the observed ranks are conceptualized as an impoverished reflection of an underlying continuous scale, and inference concerns the parameters that govern the latent representation. We apply this generic data-augmentation method to obtain Bayes factors for three popular rank-based tests: the rank sum test, the signed rank test, and Spearman's $\rho_s$. 
\end{abstract}

\noindent%
\vfill

\section{Introduction}
\label{sec:intro}
Rank-based statistical procedures offer a range of advantages. First, they are robust to outliers and to violations of distributional assumptions. Second, they are invariant under monotonic transformations, which is desirable when interest concerns a hypothesized concept (e.g., rat intelligence) whose relation to the measurement scale is only weakly specified (e.g., brain volume or log brain volume could be used as a predictor; without a process model that specifies how brain physiology translates to rat intelligence, neither choice is privileged). Third, many data sets are inherently ordinal (e.g., Likert scales, where survey participants are asked to indicate their opinion on, say, a 7-point scale ranging from `disagree completely' to `agree completely'). Finally, rank-based procedures perform better than their fully parametric counterparts when assumptions are violated, with little loss of efficiency when the assumptions do hold
\citep{hollander1973}.

Prominent rank-based tests include the Mann-Whitney-Wilcoxon rank sum test (i.e., the rank-based equivalent of the two-sample $t$-test), the Wilcoxon signed rank test (i.e., the rank-based equivalent of the paired sample $t$-test), and Spearman's $\rho_s$ (i.e., a rank-based equivalent of the Pearson correlation coefficient). These ordinal tests were developed within the frequentist statistical paradigm, and Bayesian analogues through Bayes factor hypothesis testing have, to the best of our knowledge, not yet been proposed. We speculate that the main challenge in the development of Bayesian  hypothesis tests for ordinal data is the lack of a  straightforward likelihood function. As stated by \citet[pp. 178-179]{Jeffreys1939} for the case of Spearman's $\rho_s$: 

\begin{quotation}
	``The rank correlation, while certainly useful in practice, is difficult to interpret. It is an estimate, but what is it an estimate of? That is, it is calculated from the observations, but a function of the observations has no relevance beyond the observations unless it is an estimate of a parameter in some law. Now what can this law be? [...] the interpretation is not clear.''
\end{quotation}

This difficulty can be overcome by postulating a latent, normally distributed level for the observed data (i.e., data augmentation). In other words, the rank data are conceptualized to be an impoverished reflection of richer latent data that are governed by a specific likelihood function. The latent normal distribution was chosen for computational convenience and ease of interpretation. This general procedure is widely known as data augmentation \citep{tanner1987, albert1993}, and Bayesian inference for the parameters of interest (e.g., a location difference parameter $\delta$ or an association parameter $\rho$) can be achieved using Markov chain Monte Carlo (MCMC). 

Below we first outline the general framework and then develop Bayesian counterparts for three popular frequentist rank-based procedures: the rank sum test, the signed rank test, and Spearman's rank correlation. Each of these developed Bayesian tests is accompanied by a simulation study that assesses the behavior of the test and a data example that highlights the desirable properties of rank-based inference.

\section{General Methodology}
\subsection{Latent Normal Models}
Latent normal models were first introduced by \cite{Pearson1900} as a means of modeling data from a $ 2 \times 2$ cross-classification table. The method was later extended by \cite{Pearson1922} to accommodate $r \times s$ tables. Instead of modeling the count data directly for the $2\times2$ case, Pearson assumed a latent bivariate normal level with certain governing parameters. In the case of cross-classification tables, the governing parameter is the \textit{polychoric correlation coefficient} (PCC) and refers to Pearson's correlation on the bivariate, latent normal level. 

A maximum likelihood estimator for the PCC was developed by \cite{Olssen1979} and \cite{Olssen1982}, and a Bayesian framework for the PCC was later introduced by \cite{Albert1992}. This idea was extended by \citet{Pettitt1982} to rank likelihood models, where the latent boundaries are not estimated but determined directly by the latent scores  \citep[see also][]{hoff2007, hoff2009}. 
For the two-sample location problem, a similar approach has been discussed by \cite{savage1956} and \cite{brooks1974,brooks1978}, where a continuous distribution is assumed to be underlying the observed data. Further models for ordinal data are given in \cite{mallows_nonnull_1957}, \cite{fligner_distance_1986}, \cite{fligner_multistage_1988}, 
and \cite{marden_analyzing_1995}.
However, these methods omit Bayesian hypothesis testing through Bayes factors and/or lack a straightforward interpretation of the model parameters.

In general, the latent normal methodology allows one to transform ordinal problems to parametric problems. The resulting models that are discussed here have a data-generating process, are governed by easily interpretable parameters on the latent level, and enable Bayes factor hypothesis testing. A detailed sampling algorithm of the general methodology is presented in the next section.

\subsection{Posterior Distribution and Bayes Factor}
Using Bayes' rule, the joint posterior of the model parameters $\theta$ and latent normal values (i.e., $(z^x \text{, }z^y)$), given the ranked data (i.e., $(x\text{, }y)$), can be decomposed as follows:
\begin{equation}
P(z^x\text{, } z^y\text{, } \theta \mid x\text{, }y) \propto P(x\text{, }y \mid z^x\text{, } z^y) \times P(z^x\text{, } z^y\mid\theta) \times  P(\theta).
\label{generalLatentNormalPost}
\end{equation}
In the rank-based context, the likelihood refers to the marginal of $P(x\text{, }y \mid z^x\text{, }z^y)P(z^x\text{, }z^y \mid \theta)$ with respect to the augmented variables $z^x$ and $z^y$. From a generative perspective, parameters $\theta$ produce latent normal data $z^x$ and $z^y$, and these in turn yield ordinal data $x$ and $y$. 

The first factor in the right-hand side of Equation \ref{generalLatentNormalPost}, $P(x\text{, } y  \mid z^x\text{, } z^y)$, consists of a set of indicator functions, presented below, that map the observed ranks to latent scores, such that the ordinal information (i.e., the ranking function) is preserved. This is similar to the approach of \cite{Albert1992b} and \cite{albert1993}, who sampled latent scores for binary or polytomous response data from a normal distribution that was truncated with respect to the ordinal information of the data. 
Consequently, across the MCMC iterations the ordinal information in the latent values remains constant and identical to that in the original data.
For the latent value $z^x_i$, this means that its range is truncated by the lower and upper thresholds that are respectively defined as:
\begin{equation}
a^x_i = \max\limits_{j:\text{ } x_j < x_i}\left({z^x_j}\right) 
\label{minthresh}
\end{equation}
\begin{equation}
b^x_i = \min\limits_{j:\text{ } x_j > x_i}\left({z^x_j}\right).
\label{maxthresh}
\end{equation}
For example, suppose that on a particular MCMC iteration we wish to augment the observed ordinal value $x_i$ to a latent $z_i^x$; on the latent scale, the lower threshold $a_i^x$ is given by the maximum latent value associated with all $x$ lower than $x_i$, whereas the upper threshold $b_i^x$ is determined by the minimum latent value associated with all $x$ higher than $x_i$. In order to remedy the high degree of autocorrelation that data augmentation can induce \citep{vandyk2001}, we included an additive decorrelating step documented by \cite{liu2000} and \cite{morey2008}. 

The second factor in the right-hand side of Equation \ref{generalLatentNormalPost}, $P(z^x\text{, } z^y \mid \theta )$, is the bivariate normal distribution of the latent scores given the model parameters $\theta$:
\begin{equation}
\large\begin{pmatrix}Z^X\\
Z^Y
\end{pmatrix} 
\sim  \mathcal{N}
\large\begin{bmatrix}
\vspace{-0mm}
\begin{pmatrix}
\mu_{z_x}\\
\mu_{z_y}
\end{pmatrix} ,&
\begin{pmatrix}
\sigma_{z_x} & \rho_{z_x z_y}\\
\rho_{z_x z_y} & \sigma_{z_y}\\
\end{pmatrix}
\end{bmatrix}.\\ ~\newline
\label{bivTruncNorm}
\end{equation}
In the case of analyzing the difference in location parameters, the term $\rho_{z_x z_y}$ is set to 0. In the case of analyzing the correlation, the terms  $\mu_{z_x}$ and  $\mu_{z_y}$ are set to 0.

Finally, the third factor in the right-hand side of Equation \ref{generalLatentNormalPost}, $P(\theta )$, refers to the prior distributions for the model parameters. 

After obtaining the joint posterior distribution for $\theta$ by MCMC sampling, we can either focus on \textit{estimation} and present the marginal posterior distribution for the parameter of interest, or we can conduct a Bayes factor \textit{hypothesis test} and compare the predictive performance of a point-null hypothesis $\mathcal{H}_0$ (in which the parameter of interest is fixed at a predefined value $\theta_0$) against that of an alternative hypothesis $\mathcal{H}_1$ (in which $\theta$ is free to vary; \citealp{KassRaftery1995,Jeffreys1939,lyetal2016}). The Bayes factor can be interpreted as a predictive updating factor, that is, degree to which the observed data drive a change from prior to posterior odds for the hypothesis of interest:
\begin{align}
\label{eq:testingBF}
\underbrace{ \frac{p(\mathcal{H}_1)}{p(\mathcal{H}_0)}}_{\substack{\text{Prior odds}} }
\,\,\,\,\times \,\,\,\,
\underbrace{ \frac{p(\text{data}\mid \mathcal{H}_1)}{p(\text{data} \mid  \mathcal{H}_0)}}_{\substack{\text{Bayes factor}_{10} } } \,\,\,\, = \,\,\,\, \underbrace{ \frac{p(\mathcal{H}_1  \mid \text{data})}{p(\mathcal{H}_0  \mid \text{data})}}_{\substack{\text{Posterior odds}} } 
\end{align}

For nested models, the Bayes factor be easily obtained using the Savage-Dickey density ratio \citep*{DickeyLientz1970,wagenmakers_bayesian_2010}, that is, the ratio of the posterior and prior ordinate for the parameter of interest $\theta$, under $\mathcal{H}_1$, evaluated at the point $\theta_0$ specified under $\mathcal{H}_0$:
\begin{equation}
\text{BF}_{10} = \frac{p (\theta_0 \mid \mathcal{H}_0)}{p(\theta_0 \mid \text{data}, \mathcal{H}_1)}.
\end{equation}

\section{Case 1: Wilcoxon Rank Sum Test}
\subsection{Background}
The ordinal counterpart to the two-sample $t$-test is known as the Wilcoxon rank sum test (or as the Mann-Whitney-Wilcoxon U test). It was introduced by \cite{wilcoxon_1945} and further developed by \cite{mann1947}, who worked out the statistical properties of the test. Let $x = (x_1,...,x_{n_1})$ and $y = (y_1,...,y_{n_2})$ be two data vectors that contain measurements of $n_1$ and $n_2$ units, respectively. The aggregated ranks $r^x, r^y$ (i.e., the ranking of $x$ and $y$ together) are defined as:
\[r^x_i = \text{rank of } x_i \text{ among } (x_1, \dots, x_{n_1}, y_1 \dots y_{n_2} ),\]
\[r^y_i = \text{rank of } y_i \text{ among } (x_1, \dots, x_{n_1}, y_1 \dots y_{n_2} ).\]
The test statistic $U$ is then given by summing over either $r^x$ or $r^y$, and subtracting $\frac{n_x (n_x + 1)}{2}$ or $\frac{n_y (n_y + 1)}{2}$, respectively. In order to test for a difference between the two groups, the observed value of $U$ can be compared to the value of $U$ that corresponds to no difference. This point of testing is defined as $\frac{n_1n_2}{2}$. 

To illustrate the procedure, consider the following hypothetical example. In the movie review section of a newspaper, three action movies and three comedy movies are each assigned a star rating between $0$ and $5$. Let $X = (4, 3, 1) $ be the star ratings for the action movies, and let $Y= (2, 3, 5) $  be the star ratings for the comedy movies. The corresponding aggregated ranks are $R^x= (5, 3.5, 1) $ and $R^y= (2, 3.5, 6) $. The test statistic $U$ is then obtained by summing over either $R^x$ or $R^y$, and subtracting $ \frac{3 (3 + 1)}{2} = 6$, yielding $3.5$ or $5.5$, respectively. Either of these values can then be compared to the null point which is equal to $\frac{n_1n_2}{2} = 4.5$.

An often used standardized effect size for $U$ is the rank-biserial correlation, denoted $\rho_{rb}$, which is the correlation coefficient used as a measure of association between a nominal dichotomous variable and an ordinal variable. The transformation is as follows:
\begin{equation}
\rho_{rb} = 1 - \frac{2U}{n_1 n_2}.
\end{equation}
The rank-biserial correlation can also be expressed as the difference between the proportion of data pairs where $x_i > y_j$ versus $x_i < y_j$ \citep{cureton1956,kerby2014}:
\begin{equation}
\rho_{rb} = \frac{\sum_{i=1}^{n_1} \sum_{j=1}^{n_2} Q(x_i - y_j) }{n_1 n_2},
\end{equation}
where $Q(d_i)$ is the sign indicator function defined as
\begin{equation}
\label{signindicator}
Q(d_i) =
\begin{cases}
-1& \text{if } d_i < 0 \\
+1& \text{if } d_i > 0 \\
\end{cases}.
\end{equation}
This provides an intuitive interpretation of the test procedure: each data point in $x$ is compared to each data point in $y$ and scored $-1$ or $1$ if it is lower or higher, respectively. In the movie ratings data example, there are three pairs for which $x_i > y_j$, five pairs for which $x_i < y_j$, and one pair for which $x_i = y_j$, yielding an observed rank-biserial correlation coefficient of $\frac{3 - 5}{9} = -0.22$, which is an indication that comedy movies receive slightly more positive reviews.

When the distributional assumptions are met, the rank sum test performs similarly to the parametric two-sample $t$-test in terms of Pitman asymptotic relative efficiency (ARE), that is, the ratio of the number of observations necessary to achieve the same level of power \citep{Lehmann_1999}.\footnote{The formula for calculating the ARE is $ \frac{N_V(\alpha, \beta, \theta)}{N_T(\alpha, \beta, \theta)}$, where $N_V(\alpha, \beta, \theta)$ is the sample size necessary for test $V$ to attain a power $\beta$ under the level $\alpha$ and alternative parameter value $\theta$. This is analogous for $N_T(\alpha, \beta, \theta)$ and test $T$.} Specifically, the rank sum test has a Pitman ARE of $\nicefrac{3}{\pi} \approx 0.955$ when the data are normally distributed \citep{hodges1956,Lehmann_1975}. Thus, even when the distributional assumption of the $t$-test holds, the rank sum test is still relatively efficient for large data sets. The ARE increases as the data distribution grows more heavy-tailed, with a maximum value of infinity. In addition, results for other distributions include the logistic distribution (ARE = $\nicefrac{\pi^2}{9} \approx 1.097$), the Laplace distribution (ARE = $1.5$), and the exponential distribution (ARE = $3$); these ARE values $ > 1$ indicate that the rank-based test outperforms the $t$-test \citep{vandervaart2000}. 

\subsection{Sampling Algorithm}

\begin{figure}[h!]
	\centering
	\centerline{ 
		\includegraphics[trim=1cm 0cm 0cm 0cm,clip=true,scale=0.8]{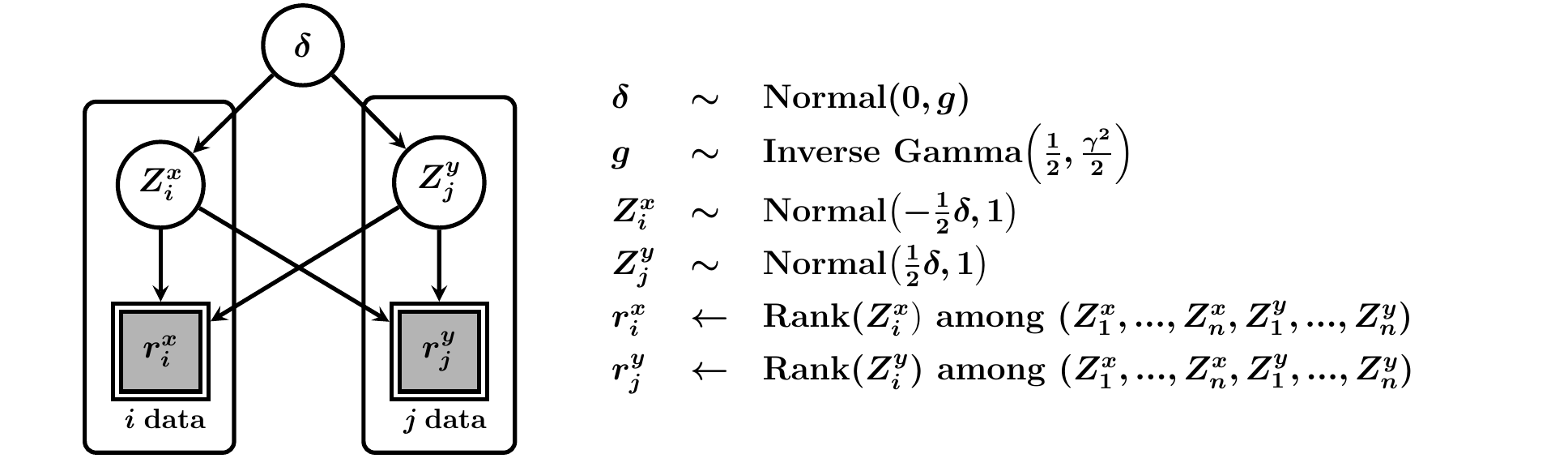}}
	\caption{The graphical model underlying the Bayesian rank sum test. The latent, continuous scores are denoted by $Z_i^x$ and $Z_i^y$, and their manifest rank values are denoted by $x_i$ and $y_j$. The latent scores are assumed to follow a normal distribution governed by the parameter $\delta$. This parameter is assigned a Cauchy prior distribution, which for computational convenience is reparameterized to a normal distribution with variance $g$ (which is then assigned an inverse gamma distribution).}
	\label{fig:twoSampleGraphModel}
\end{figure}

The Bayesian data augmentation algorithm for the rank sum test follows the graphical model outlined in Figure \ref{fig:twoSampleGraphModel}. The ordinal information contained in the aggregated ranking constrains the corresponding values for the latent normal parameters $Z^x$ and $Z^y$ to lie within certain intervals (i.e., the ordinal information imposes truncation).
The parameter of interest here is the effect size $\delta$, the difference in location of the distributions for $Z^x$ and $Z^y$. We follow \citet{jeffreys1961} and assign $\delta$ a Cauchy prior with scale parameter $\gamma$. For computational simplicity, this prior is implemented as a normal distribution with an inverse gamma prior on the variance, where the shape parameter is set to $0.5$ and the scale parameter is set to $\nicefrac{\gamma^2}{2}$ \citep*{Liang_2008,rouderetal2009ttest}. The difference with earlier work is that we set $\sigma$ to $1$, as the rank data contain no information about the variance and the inclusion of $\sigma$ in the sampling algorithm becomes redundant.

In order to sample from the posterior distributions of $\delta$, $Z^x$ and $Z^y$, we used Gibbs sampling \citep{Geman1984}. Specifically, the sampling algorithm for the latent $\delta$ is as follows, at sampling time point $s$:
\begin{enumerate}
	\item For each $i$ in $(1\text{, } \dots\text{, } n_x) $, sample $Z^x_i$ from a truncated normal distribution, where the lower threshold is $a^x_i$ given in \eqref{minthresh} and the upper threshold is $b^x_i$ given in \eqref{maxthresh}:
	
	\[ (Z_i^x \mid z_{i'}^{x}\text{, } z_i^y, \delta) \sim \mathcal{N}_{\left(a_i^x\text{, } b_i^x\right)}\left(-\tfrac{1}{2}\delta \text{, } 1 \right), \]
	where the subscripts of $\mathcal{N}$ indicate the interval that is sampled from.
	\item  For each $i$ in $(1\text{, } \dots\text{, } n_y) $, the sampling procedure for $Z^y_i$ is analogous to step 1, with
	\[ (Z_i^y \mid z_{i'}^{y}\text{, } z_i^x\text{, } \delta) \sim \mathcal{N}_{\left(a_i^y\text{, } b_i^y\right)}\left(\tfrac{1}{2}\delta\text{, } 1 \right). \] 
	\item  Sample $\delta$ from 
	\[(\delta \mid z^x\text{, }z^y\text{, } g) \sim \mathcal{N}(\mu_\delta, \sigma_\delta), \]
	where
	\[ \mu_\delta  = \frac{2  g (n_y\overbar{z^y} - n_x\overbar{z^x}) }{g(n_x + n_y) + 4} \]
	\[ \sigma^2_\delta =  \frac{4g}{g(n_x + n_y)+4} .\]
	\item  Sample $g$ from 
	\[(G \mid \delta) \sim \text{Inverse Gamma}\left( 1 \text{, } 
	\frac{\delta^2 + \gamma^2 }{2} \right)\text{, } \]
	where $\gamma$ determines the scale (i.e., width) of the Cauchy prior on $\delta$. 
	
\end{enumerate}
Repeating the algorithm a sufficient number of times yields samples from the posterior distributions of $Z^x, Z^y,$ and $\delta$. The posterior distribution of $\delta$ can then be used to obtain a Bayes factor through the Savage-Dickey density ratio.

\subsection{Simulation Study}
In order to provide insight into the behavior of the inferential framework, a simulation study was performed. 
For three values of difference in location parameters, $\delta$ (0, 0.5, 1.5), and three values of $n$ (10, 20, 50), $1{,}000$ data sets were generated under various distributions: skew-normal, Cauchy, logistic, and uniform distributions. In one scenario, both groups have the same distributional shape (e.g., both follow a logistic distribution), and in a second scenario, one group follows the normal distribution and one group follows one of the aforementioned distributions. 

\begin{figure}[h!]
	\centering
	\centerline{ 
		\includegraphics[trim=0cm 0cm 0cm 0cm,clip=true,scale=0.6]{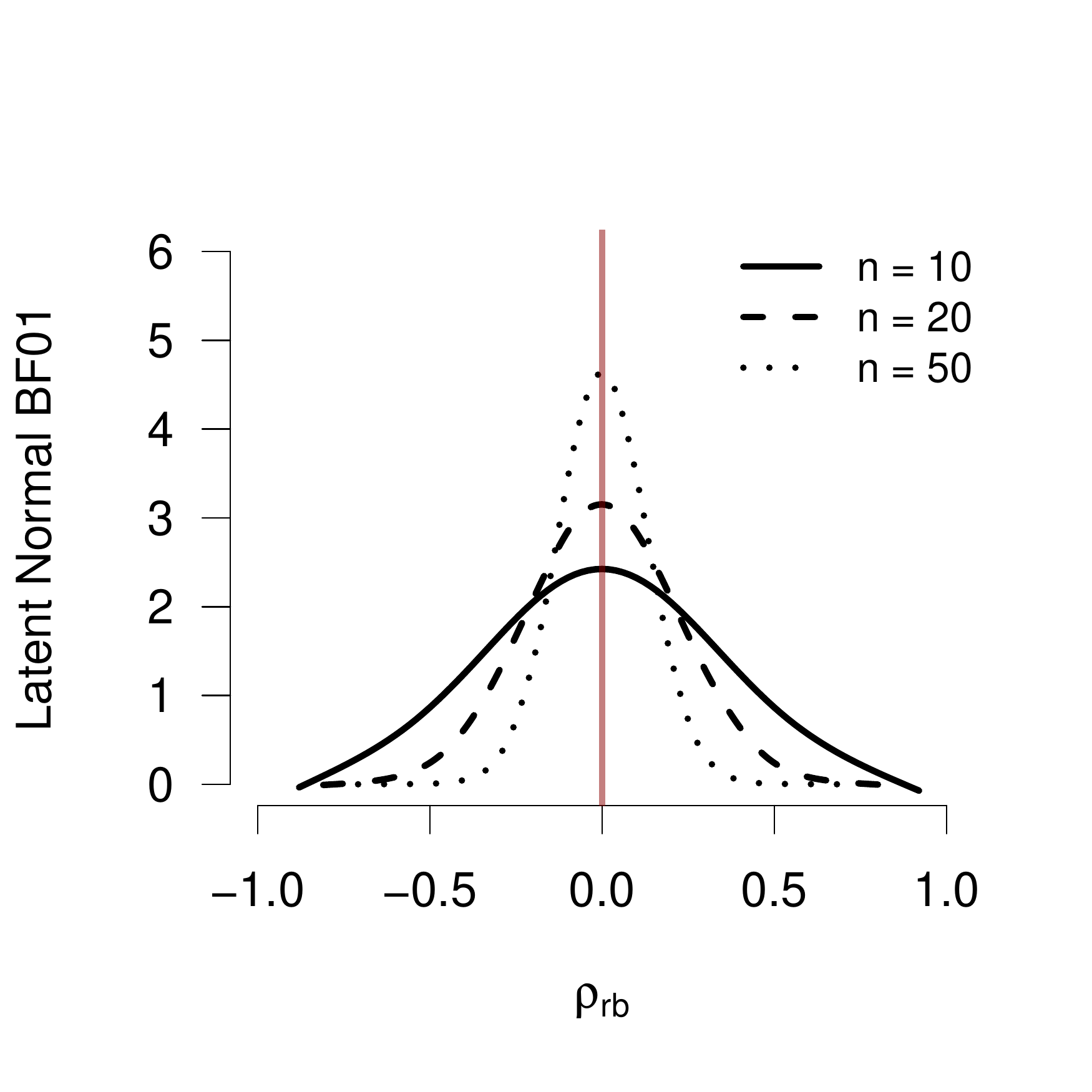}}
	\caption{The relationship between the latent normal Bayes factor and the observed rank-based test statistic is illustrated for logistic data. Because $U$ is dependent on $n$, the rank biserial correlation coefficient is plotted on the x-axis instead of $U$. The relationship is clearly defined, and maximum evidence in favor of $\mathcal{H}_0$ is attained when $\rho_{rb} = 0$. The further $\rho_{rb}$ deviates from 0, the stronger the evidence in favor of $\mathcal{H}_1$ becomes. The lines depict smoothing splines fitted to the observed Bayes factors.}
	\label{fig:rankSumBF_W}
\end{figure}

First, the relationship between the observed rank statistic $U$ and the latent normal Bayes factor was analyzed. Figure \ref{fig:rankSumBF_W} illustrates this relationship, fitted with a cubic smoothing spline \citep{chambers1992}, for two logistic distributions ($\alpha = 20$). 
To show results for multiple values of $n$ in one figure, the rank biserial correlation coefficient $\rho_{rb}$ is plotted instead of $U$. The figure shows a clear relationship, where the Bayes factor favors $\mathcal{H}_0$ in cases where the observed statistic is close to the null-point of no difference between groups ($\rho_{rb} = 0$). This relationship grows more decisive as $n$ increases. The results are highly similar for the other distributions that were considered (see the online supplementary material at \url{https://osf.io/gny35/} for the results of these scenarios). Since both metrics, $\rho_{rb}$ and $\text{BF}_{01}$, depend solely on the ordinal information in the data, the observed relationship is not surprising. This result highlights and illustrates the robustness of the latent normal Bayes factor to violations of the assumptions of the parametric test: it illustrates the same robustness as the traditional $W$ test statistic.

\begin{figure}[h!]
	\centering
	\centerline{ 
		\includegraphics[trim=0cm 0cm 0cm 0cm,clip=true,scale=0.6]{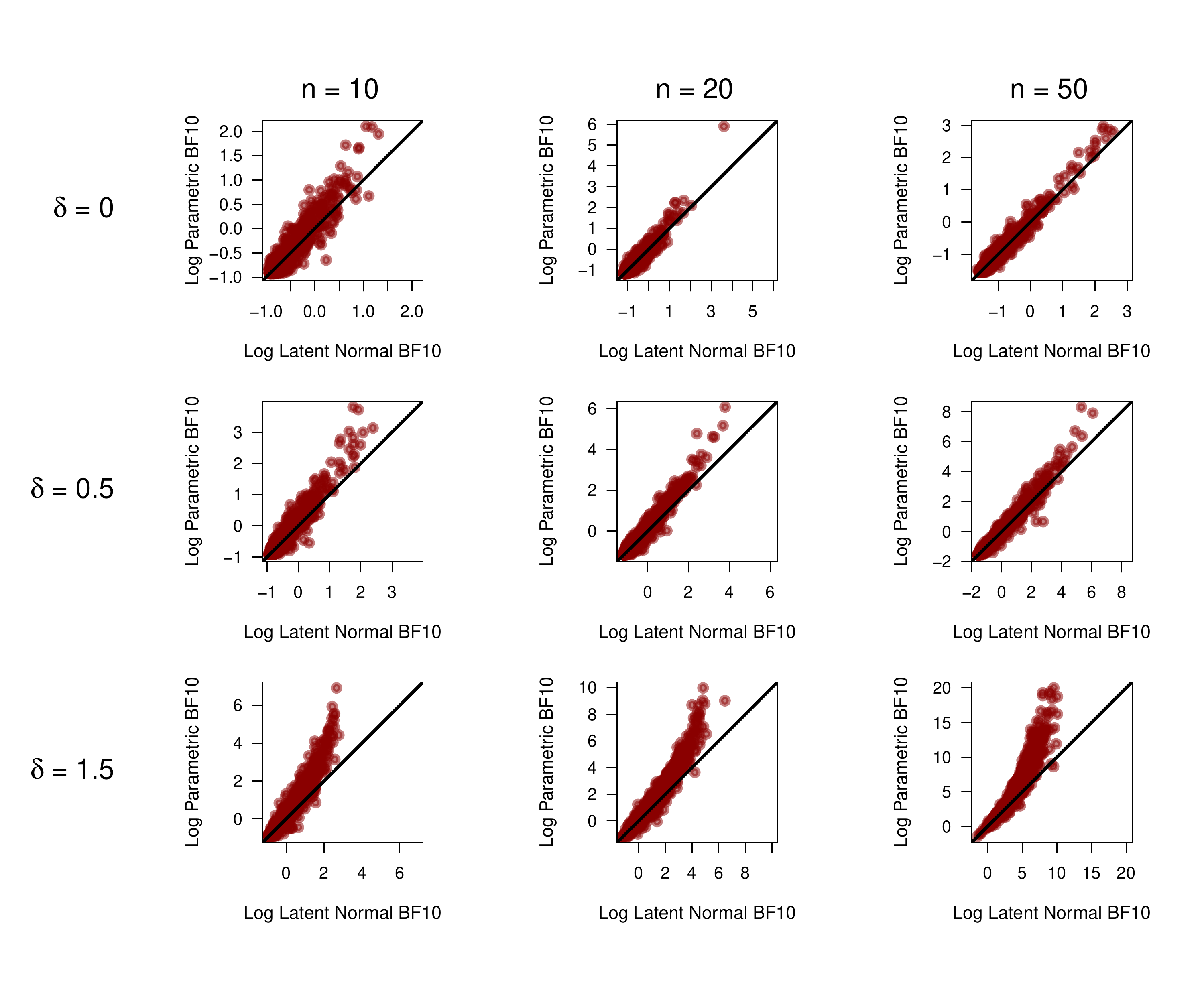}}
	\caption{For all combinations of difference in location parameters $\delta$, and $n$, the relationship between the latent normal Bayes factor and the parametric Bayes factor is shown for logistic data. The black lines indicate the point of equivalence. The two Bayes factors are generally in agreement, as suggested by the ARE results in \cite{vandervaart2000}. }
	\label{fig:rankSumBFcomp}
\end{figure}

Second, the relationship between the latent normal Bayes factor and the parametric Bayes factor \citep{rouderetal2009ttest} was analyzed. Figure \ref{fig:rankSumBFcomp} illustrates this relationship for all values of $n$ and $\delta$ that were used, again in the scenario with two logistic distributions. Generally, the two Bayes factors are in agreement. In cases where $\delta$ deviates from 0,  the parametric Bayes factor becomes more decisive compared to the latent normal Bayes factor.
For distributions of data that violate the assumptions of the parametric test, such as the Cauchy distribution, the relationship between the two Bayes factors is notably less defined. In this case, the parametric test greatly overestimates the variance and is no longer able to detect differences in location parameters  (see the supplementary material), whereas the latent normal Bayes factor is unaffected by this.

\subsection{Data Example}
\citet{cortez2008} gathered data from $395$ students concerning their math performance (scored between $1$ and $20$) and their level of alcohol intake (self-rated on a Likert scale between $1$ and $5$). Students passed the course if they scored $\geq 10$, and we will test whether students who failed the course ($n_1 = 130$) had a higher self-reported alcohol intake than their peers who passed ($n_2 = 265$).

\begin{figure}[h!]
	\centering
	\centerline{ 
		\includegraphics[trim=0cm 0cm 0cm 0cm,clip=true,scale=0.6]{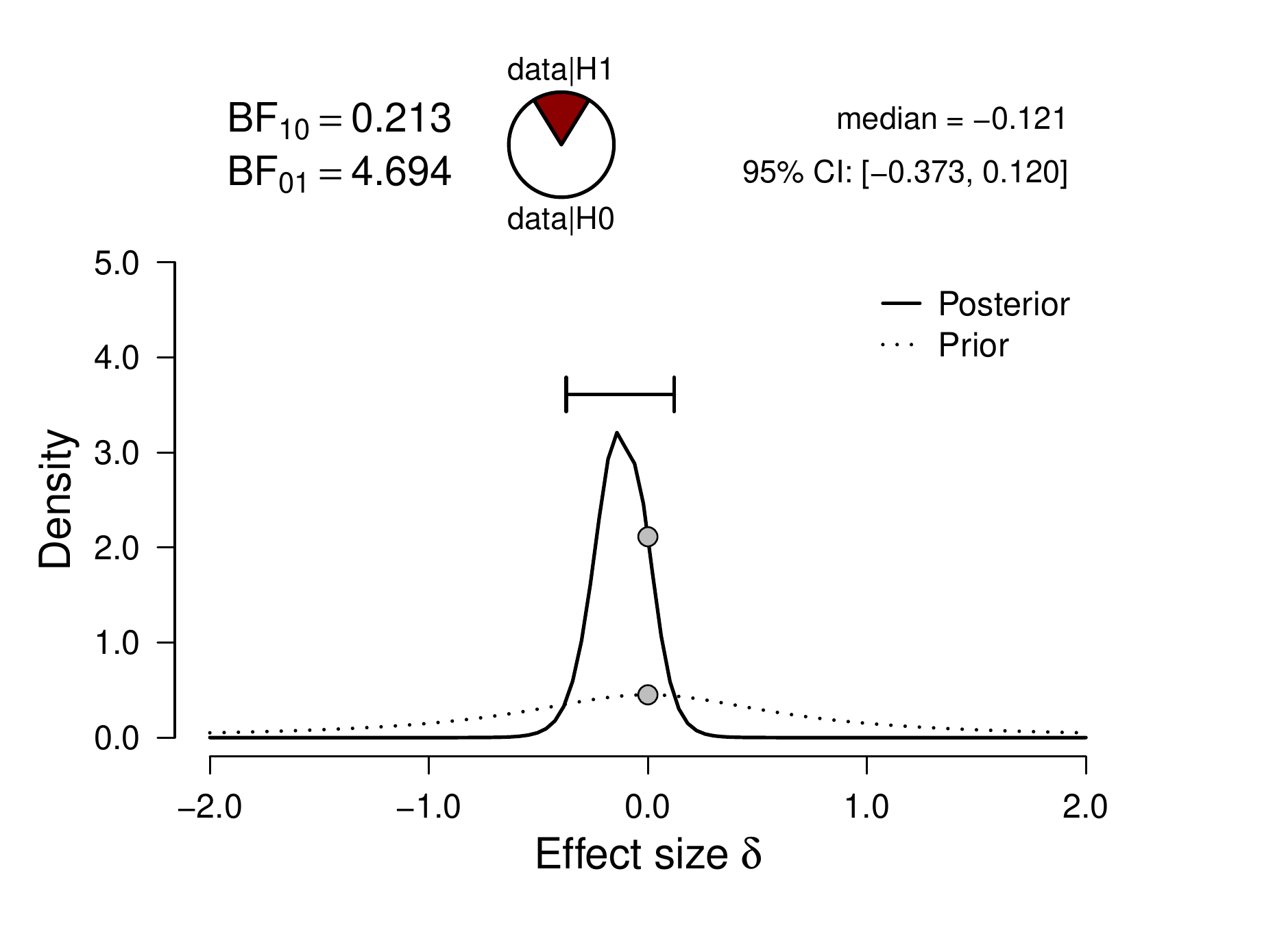}}
	\caption{Do students who flunk a math course report drinking more alcohol? Results for the Bayesian rank sum test as applied to the data set from \citet{cortez2008}. The dashed line indicates the Cauchy prior with scale  $\nicefrac{1}{\sqrt{2}}$. The solid line indicates the posterior distribution. The two grey dots indicate the prior and posterior ordinate at the point under test, in this case $\delta = 0$. The ratio of the ordinates gives the Bayes factor.}
	\label{fig:twoSampleDataExample}
\end{figure}

As alcohol intake was measured on a Likert scale, the data contain many ties and show extreme non-normality. These properties make this data set particularly suitable for the latent-normal rank sum test.
The hypotheses can be specified as follows:
\[ \mathcal{H}_0 : \delta = 0, \]  
\[ \mathcal{H}_1 : \delta \sim \text{Cauchy}\left(0\text{, }\frac{1}{\sqrt{2}}\right). \]
The null hypothesis posits that alcohol intake does not differ between the students who passed the course and those who failed. The alternative hypothesis posits the presence of an effect and assigns effect size a Cauchy distribution with scale parameter set to $\nicefrac{1}{\sqrt{2}}$, as advocated by \cite{MoreyRouderBayesFactorPackage}. 
Figure \ref{fig:twoSampleDataExample} shows the resulting posterior distribution for $\delta$ under $ \mathcal{H}_1 $ and the associated Bayes factor. The posterior median for $\delta$ equals $-0.121$, with a $95\%$ credible interval that ranges from $-0.373$ to $0.120$. The corresponding Bayes factor indicates that the data are about $4.694$ times more likely under $ \mathcal{H}_0$ than under $\mathcal{H}_1$, indicating moderate evidence against the hypothesis that self-reported alcohol intake differentiates between students who did and who did not pass the math exam.

\section{Case 2: Wilcoxon Signed Rank Test}
\subsection{Background}
The rank-based counterpart to the paired samples $t$-test was proposed by \cite{wilcoxon_1945}, who termed it the \textit{signed rank test}. The test procedure involves taking the difference scores between the two samples under consideration and ranking the absolute values. 
The procedure may also be applied to one-sample scenarios by ranking the differences between the observed sample and the point of testing. 
These ranks are then multiplied by the sign of the respective difference scores and summed to produce the test statistic $W$. For the paired samples signed rank test, let $x = (x_1\text{, }...\text{, }x_n)$ and $y = (y_1,...,y_n)$ be two data vectors each containing measurements of the same $n$ units, and let $d = (d_1\text{, }...\text{, }d_n)$ denote the difference scores. For the one-sample signed rank test, this process is analogous, except $y$ is replaced by the test value. The test statistic is then defined as: \[W = \sum_{1}^{n}\left[ \text{rank}(\lvert d_i \rvert) \times Q(d_i)  \right],  \]
where $Q$ is the sign indicator function given in \eqref{signindicator}.

To illustrate the procedure, consider the following hypothetical data example. Three students take a math exam, graded between $0$ and $10$, before and after receiving a tutoring session. Let $X = (5, 8, 4) $ be their scores on the exam before the session, and let $Y= (6, 7, 7) $  be their scores on the exam after the session. The difference scores, the ranks of the absolute difference scores, and the sign indicator function are presented in Table \ref{small_example_onesample}. In order to have a positive test statistic indicate an increase in scores, the difference scores are defined here as $(y_i - x_i)$.
The test statistic $W$ is then calculated by summing over the product of the fourth and fifth column: $1.5 - 1.5 + 3 = 3$. This value indicates a slight increase in math scores after the tutoring session.

\begin{table}[h]
	\centering
	\begin{tabular}{ccccc}
		\vspace{2pt}
		$i$ &  $(y_i - x_i)$ & $d_i$ & $\text{rank}(\lvert d_i \rvert)$ & $Q(d_i)$ \\ \hline \\ [-10pt]
		$1$ &  $6-5$    		& $1$	& $1.5$	  & $1$   \\
		$2$ &  $7-8$    		& $-1$    	& $1.5$     & $-1$  \\
		$3$ &  $7-4$    		& $3$    & $3$	      & $1$
	\end{tabular}
	\caption{The scores, difference scores, ranks of the absolute difference scores, and the sign indicator function $Q$ for the hypothetical scenario where $X = (5, 8, 4)$ are the initial scores on a math exam and $Y = (6, 7, 7)$ are the scores on the exam after a tutoring session.}
	\label{small_example_onesample}
\end{table}

An often used standardized effect size for $W$ is the matched-pairs rank-biserial correlation, denoted $\rho_{mrb}$, which is the correlation coefficient used as a within subjects measure of association between a nominal dichotomous variable and an ordinal variable \citep{cureton1956,kerby2014}. The transformation is as follows:
\begin{equation}
\rho_{mrb} = 1 - \frac{4W}{n ( n+1)}.
\end{equation}
The matched-pairs rank-biserial correlation can also be expressed as the difference between the proportion of data pairs where $x_i > y_i$ versus $x_i < y_i$. For the grades example, there is one pair for which $x_i > y_i$, and two pairs for which $x_i < y_i$, yielding a matched-pairs rank-biserial correlation coefficient of $\frac{2-1}{3} = \frac{2}{3}$, which is an indication that the tutoring session has increased students' math ability.

The signed rank test is similar to the sign test, where the procedure is to sum over the sign indicator function. The difference here is that the output of the sign indicator function is weighted by the ranked magnitude of the absolute differences. The signed rank test has a higher ARE than the sign test: a relative efficiency of $\frac{3}{2}$ for all distributions \citep{conover1999}. 
For the one-sample scenario, the Pitman ARE of the signed rank test (compared to the fully parametric $t$-test) is similar to the ARE of the rank sum test for the unpaired two-sample scenario; for example, when the data follow a normal distribution the ARE equals $\frac{3}{\pi}$. For other distributions, especially when these are heavy-tailed, the signed rank test outperforms the $t$-test \citep{Lehmann_1999,vandervaart2000}.


\subsection{Sampling Algorithm}

\begin{figure}[h!]
	\centering
	\centerline{
		\includegraphics[trim=0.5cm 0cm 0cm 0cm,clip=true,scale=1]{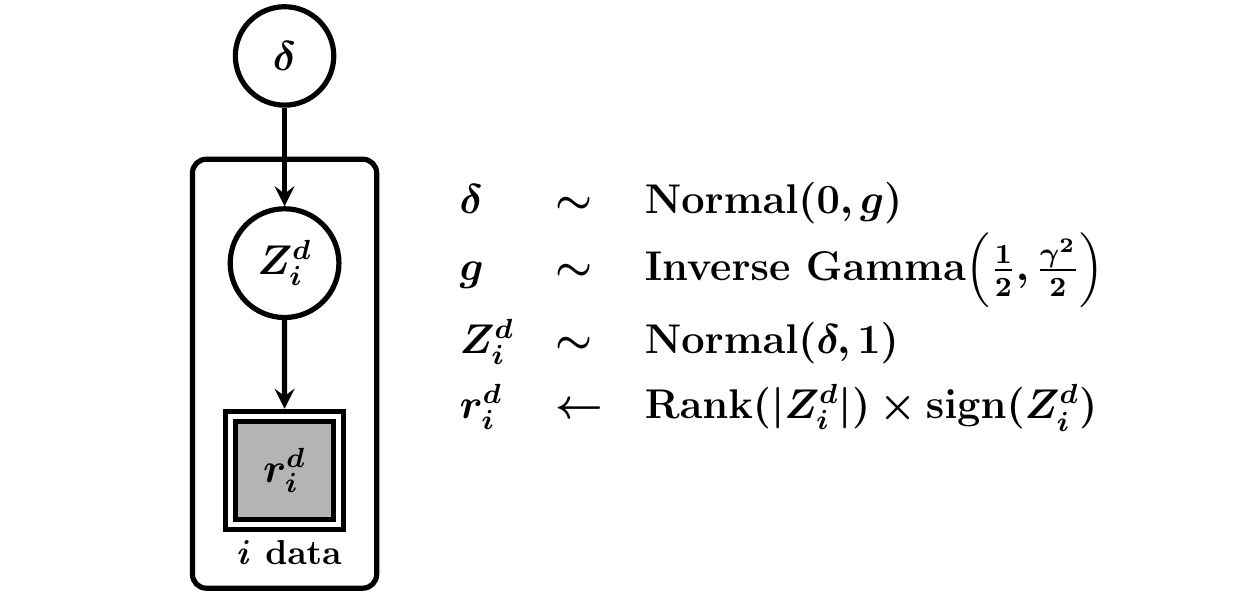}}
	\caption{The graphical model underlying the Bayesian signed rank test. The latent, continuous difference scores are denoted by $Z_i^d$, and their manifest signed rank values are denoted by $d_i$. The latent scores are assumed to follow a normal distribution governed by parameter $\delta$. This parameter is assigned a Cauchy prior distribution, which for computational convenience is reparameterized to a normal distribution with variance $g$ (which is then assigned an inverse gamma distribution).}
	\label{fig:oneSampleGraphModel}
\end{figure}

The data augmentation algorithm is similar to that of the rank sum test and is outlined in Figure \ref{fig:oneSampleGraphModel}. Here $d$ denotes the difference scores as ordinal manifestations of latent, normally distributed values $Z^d$. The parameter of interest is again the standardized location parameter $\delta$, which is assigned a Cauchy prior distribution with scale parameter $\gamma$. Similar to the rank sum sampling procedure, the variance of $Z^d$ is set to $1$, as the ranked data contain no information about the variance. The computational complexity of sampling from the posterior distribution of $\delta$ is again reduced by introducing the parameter $g$. 
The Gibbs algorithm for the data augmentation and sampling $\delta$ is as follows, at sampling time point $s$:
\begin{enumerate}
	\item For each value of $i$ in $(1\text{, } \dots \text{, } n)$, sample  $Z^d_i$ from a truncated normal distribution, where the lower threshold is $a^d_i$ given in \eqref{minthresh} and the upper threshold is $b^d_i$ given in \eqref{maxthresh}:
	\[ (Z_i^d \mid z_{i'}^{d}\text{, }\delta) \sim \mathcal{N}_{\left(a_i^d, \, b_i^d\right)}\left(\delta \text{, } 1\right) \]
	
	\item  Sample $\delta$ from 
	\[(\delta \mid z^d\text{, } g) \sim \mathcal{N} \left(\mu_\delta \text{, } \sigma^2_{\delta}\right)\text{, } \]
	where 
	\[ \mu_{\delta} = \frac{  gn \overbar{z^d} }{gn+1}\]
	\[ \sigma^2_{\delta} = \frac{g}{gn+1}\]

	\item  Sample $g$ from 
	\[(g \mid \delta) \sim \text{Inverse Gamma}\left( 1 \text{, } 
	\frac{\delta^2 + \gamma^2 }{2} \right)\text{, } \]
	where $\gamma$ determines the scale (i.e., width) of the Cauchy prior on $\delta$.

\end{enumerate}
Repeating the algorithm a sufficient number of times yields samples from the posterior distributions of $Z^d$ and $\delta$. The posterior distribution of $\delta$ can then be used to obtain a Bayes factor through the Savage-Dickey density ratio.

\subsection{Simulation Study}
Similar to the Wilcoxon rank sum test, a simulation study was performed to illustrate the behavior of the Bayesian signed rank test. For three values of difference in location parameters, $\delta$ (0, 0.5, 1.5), and three values of $n$ (10, 20, 50), $1{,}000$ data sets were generated under various distributions: skew-normal, Cauchy, logistic, and uniform distributions. In one scenario, both groups have the same distributional shape, and in a second scenario, one group follows the normal distribution and one group follows one of the aforementioned distributions. After the data were generated, the difference scores between the two groups were calculated, and used as input for the Bayesian latent normal test. 

\begin{figure}[h!]
	\centering
	\centerline{ 
		\includegraphics[trim=0cm 0cm 0cm 0cm,clip=true,scale=0.6]{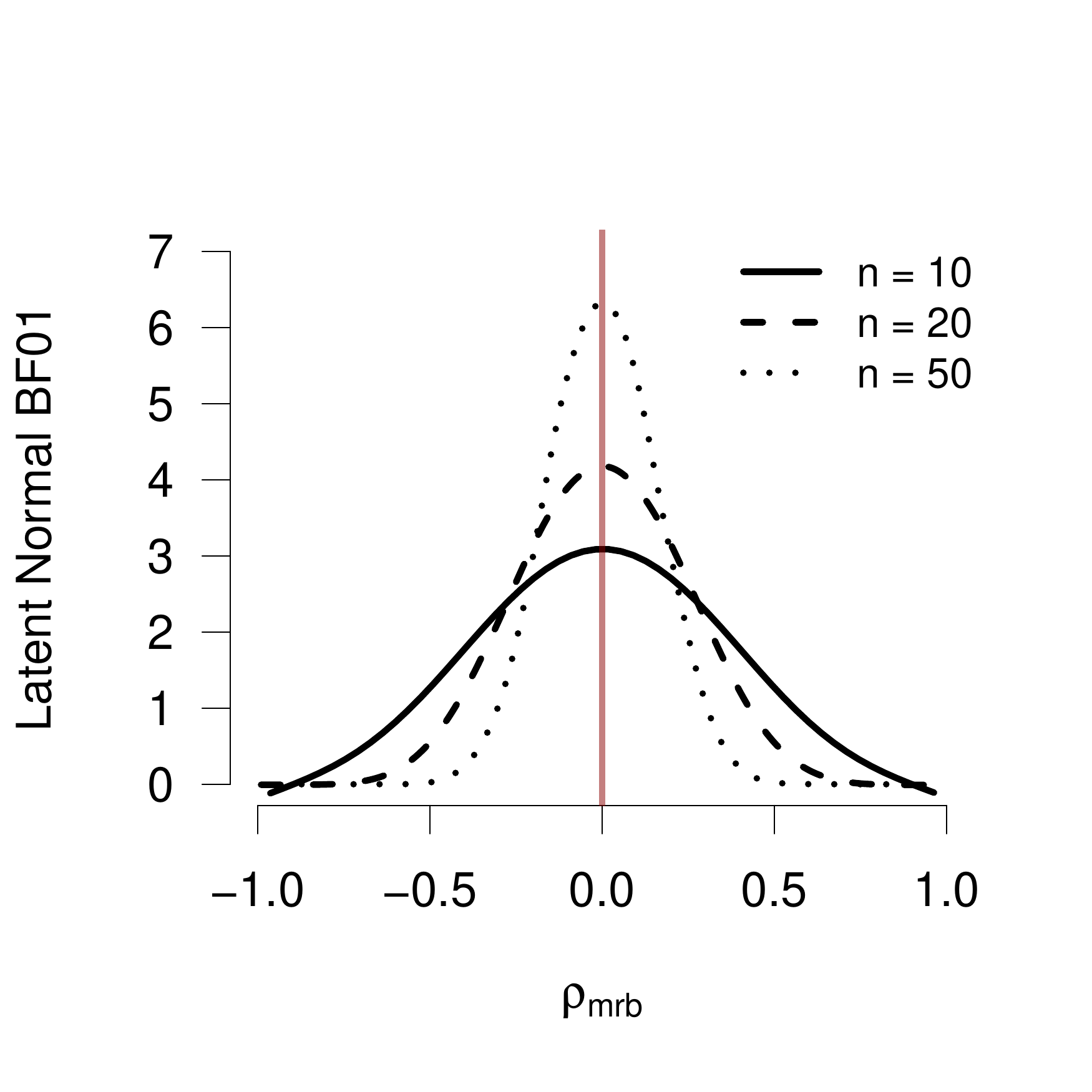}}
	\caption{The relationship between the latent normal Bayes factor and the observed rank-based test statistic is illustrated  for logistic data. Because $W$ is dependent on $n$, the matched-pairs rank-biserial correlation coefficient is plotted on the x-axis instead of $W$. The relationship is clearly defined, and maximum evidence in favor of $\mathcal{H}_0$ is attained when $\rho_{mrb} = 0$. The further $\rho_{mrb}$ deviates from 0, the stronger the evidence in favor of $\mathcal{H}_1$ becomes. The lines are smoothing splines fitted to the observed Bayes factors.}
	\label{fig:signRankBF_W}
\end{figure}

The same analyses were performed as for the Wilcoxon rank sum test. First, the relationship between the observed rank statistic $W$ and the latent normal Bayes Factor was analyzed. Figure \ref{fig:signRankBF_W} illustrates this relationship, fitted with a cubic smoothing spline \citep{chambers1992}, when the difference scores were taken for two logistic distributions. To show results for multiple values of $n$ in one figure, the matched-pairs rank-biserial correlation coefficient $\rho_{mrb}$ is plotted instead of $W$. The Bayes factor shows a clear relationship with the rank-based test statistic, where the maximum evidence in favor of $\mathcal{H}_0$ is obtained when this statistic equals $0$. Furthermore, the obtained Bayes factor grows more decisive as $n$ increases. For other distributions of the data, highly similar results were obtained (see the online supplementary material at \url{https://osf.io/gny35/} for the results of these scenarios). 

\begin{figure}[h!]
	\centering
	\centerline{ 
		\includegraphics[trim=0cm 0cm 0cm 0cm,clip=true,scale=0.6]{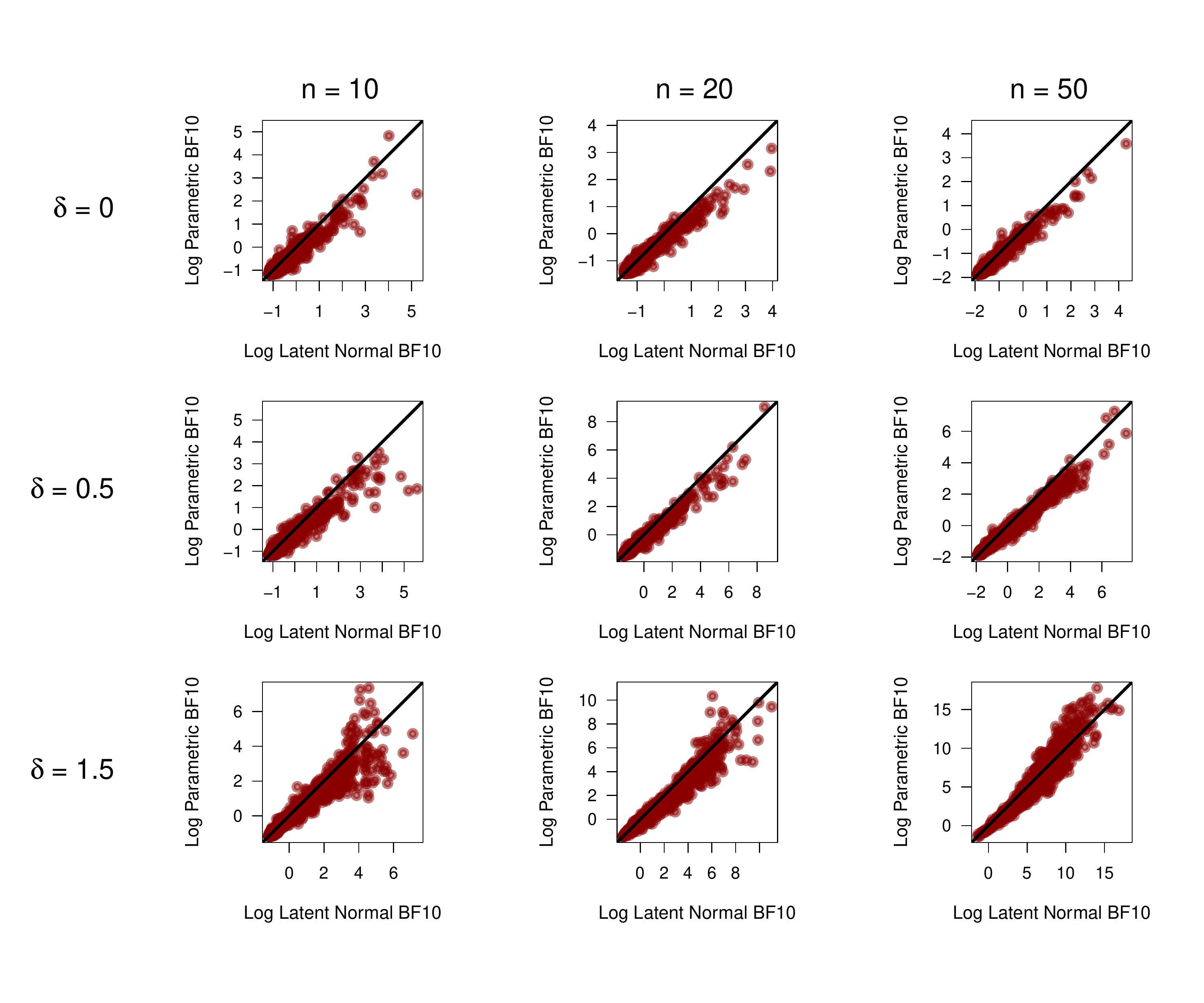}}
	\caption{For all combinations of difference in location parameters $\delta$, and $n$, the relationship between the latent normal Bayes factor and the parametric Bayes factor is shown for logistic data. The black lines indicate the point of equivalence. The two Bayes factors are generally in agreement, with the latent normal Bayes factor accumulating evidence in favor of the true model faster.}
	\label{fig:signRankBFcomp}
\end{figure}

Next to the relationship between $W$ and the latent normal Bayes factor, the relationship between the latent normal Bayes factor and the parametric Bayes factor \citep{rouderetal2009ttest} was analyzed. Figure \ref{fig:signRankBFcomp} illustrates the results for all combinations of $n$ and the difference in location parameters, $\delta$.  
The two Bayes factors are generally in agreement, with the parametric Bayes factor accumulating evidence in favor of $\mathcal{H}_1$ faster when this is the true model. The latent normal Bayes factor demonstrates more instability, due to only using the ordinal information in the data. 
For distributions of the data that violate the assumptions of the parametric test, such as the Cauchy distribution, the parametric test greatly overestimates the variance and is no longer able to detect differences in location parameters (see the supplementary material). This misspecification does not affect the latent normal Bayes factor, underscoring its robustness.

\subsection{Data Example}
\cite{thall1990some} investigated a data set obtained by D. S. Salsburg concerning the effects of the drug progabide on the occurrence of epileptic seizures. During an initial eight week baseline period, the number of epileptic seizures was recorded in a sample of $31$ epileptics. Next, the patients were given progabide, and the number of epileptic seizures was recorded for another eight weeks. In order to accommodate the discreteness and non-normality of the data, \citet{thall1990some} applied a log-transformation on the counts. 

This log-transformation has a clear impact on the outcome of a parametric Bayesian $t$-test \citep{MoreyRouderBayesFactorPackage}: $\text{BF}_{10} \approx 0.2$ for the raw data, whereas $\text{BF}_{10} \approx 2.95$ for the log-transformed data. Here we analyze the data with the signed rank test; because this test is invariant under monotonic transformations, the same inference will result regardless of whether or not the data are log-transformed.

The hypothesis specification here is similar to that of the previous setup:
\[ \mathcal{H}_0 : \delta = 0, \]  
\[ \mathcal{H}_1 : \delta \sim \text{Cauchy}\left(0\text{, }\frac{1}{\sqrt{2}}\right), \]
where the null hypothesis postulates that the effect is absent whereas the alternative hypothesis assigns effect size a Cauchy prior distribution. 

\begin{figure}[h!]
	\centering
	\centerline{ 
		\includegraphics[trim=0cm 0cm 0cm 0cm,clip=true,scale=0.6]{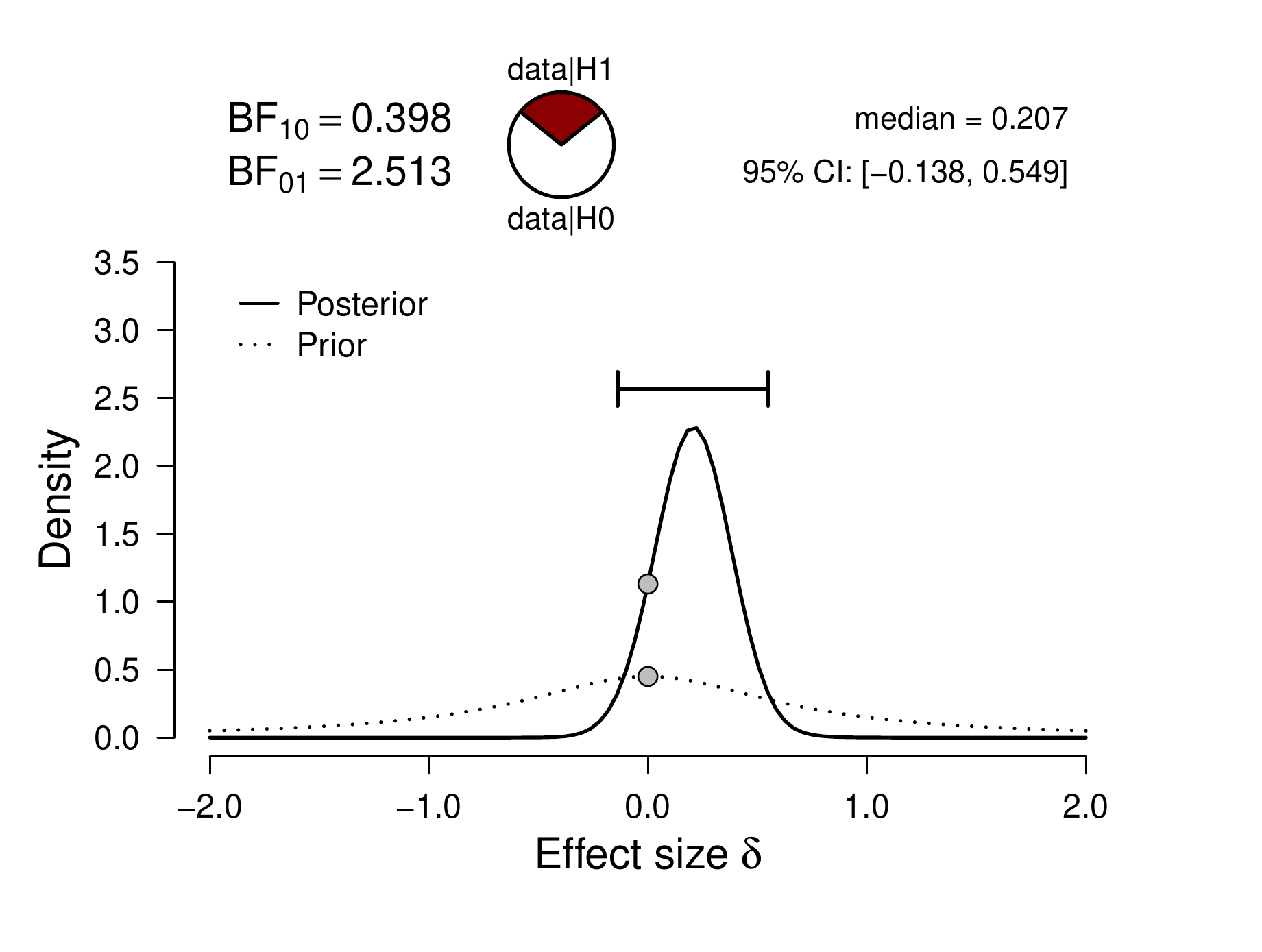}}
	\caption{Does progabide reduce the frequency of epileptic seizures? Results for the Bayesian signed rank test as applied to the data set presented in \citet{thall1990some}. The dashed line indicates the Cauchy prior with scale $\nicefrac{1}{\sqrt{2}}$. The solid line indicates the posterior distribution. The two grey dots indicate the prior and posterior ordinate at the point under test, in this case $\delta = 0$. The ratio of the ordinates gives the Bayes factor.}
	\label{fig:oneSampleData}
\end{figure}

Figure \ref{fig:oneSampleData} shows the resulting posterior distribution for $\delta$ under $ \mathcal{H}_1 $ and the associated Bayes factor. The posterior median for $\delta$ equals $0.207$, with a $95\%$ credible interval that ranges from $-0.138$ to $0.549$. The corresponding Bayes factor indicates that the data are about $2.513$ times more likely under $ \mathcal{H}_0$ than under $\mathcal{H}_1$, indicating that, for the purpose of discriminating $ \mathcal{H}_0$ from $\mathcal{H}_1$, the data are almost perfectly uninformative.

\section{Case 3: Spearman's $\rho_s$}
\subsection{Background}
\citet{spearman1904proof} introduced the rank correlation coefficient $\rho$ in order to overcome the main shortcoming of Pearson's product moment correlation, namely its inability to capture monotonic but non-linear associations between variables. Spearman's method first applies the rank transformation on the data and then computes the product-moment correlation on the ranks. Let $x = (x_1\text{, }...\text{, }x_n)$ and $y = (y_1,...,y_n)$ be two data vectors each containing measurements of the same $n$ units, and let $r^x = (r^x_1\text{, }...\text{, }r^x_n)$ and $r^y = (r^y_1\text{, }...\text{, }r^y_n)$ denote their rank-transformed values, where each value is assigned a ranking within its variable. 
This then leads to the following formula for Spearman's $\rho_s$:
\[ \rho_s = \frac{\text{Cov}_{r^x r^y }}{\sigma_{r^x}\sigma_{r^y}} .\]

\subsection{Sampling Algorithm}
\begin{figure}[h!]
	\centering
	\centerline{ 
		\includegraphics[trim=0cm 0cm 0cm 0cm,clip=true,scale=0.9]{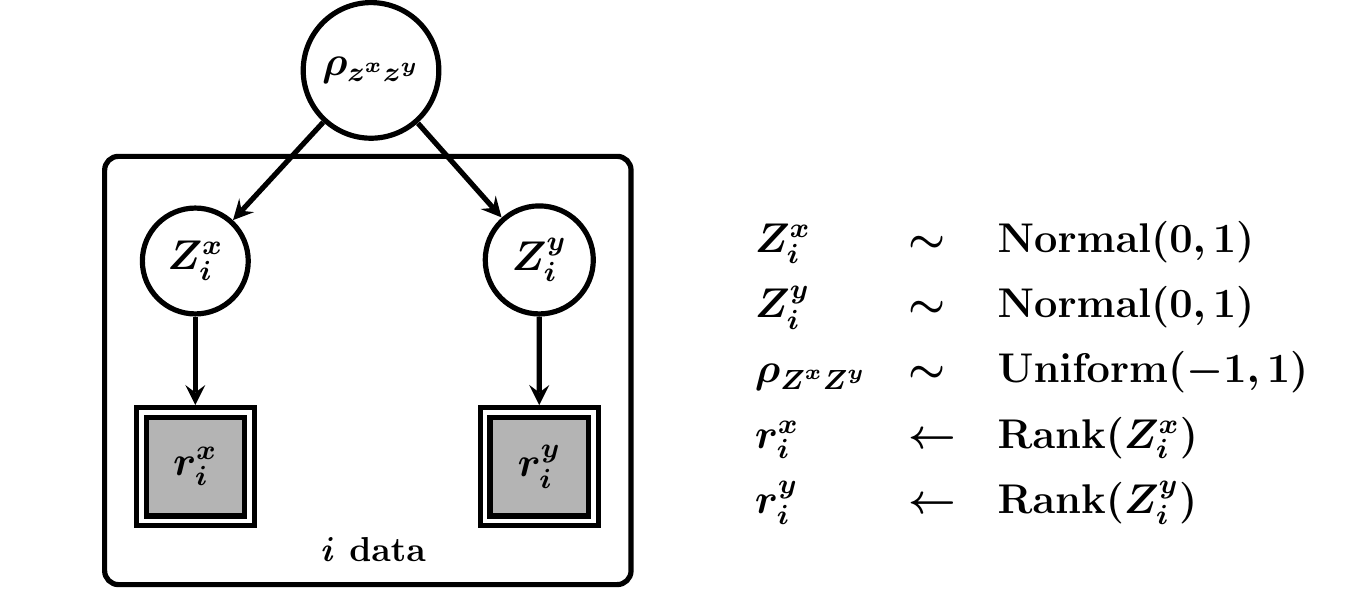}}
	\caption{The graphical model underlying the Bayesian test for Spearman's $\rho_s$. The latent, continuous scores are denoted by $Z_i^x$ and $Z_i^y$, and their manifest rank values are denoted by $r^x_i$ and $r^y_j$. The latent scores are assumed to follow a normal distribution governed by parameter $\rho$ (which is assigned a uniform prior distribution).}
	\label{fig:corGraphModel}
\end{figure}

The graphical model in Figure \ref{fig:corGraphModel} illustrates the data augmentation setup for inference on the latent correlation parameter $\rho$. The sampling method is a Metropolis-within-Gibbs algorithm, where data augmentation is conducted with a Gibbs sampling algorithm as before, but combined with a random walk Metropolis-Hastings sampling algorithm \citep{Metropolis1953,Hastings1970} to sample from the posterior distribution of $\rho$ (see also \citealp{vanDoorn2019}).

The sampling algorithm for the latent correlation is as follows, at sampling time point $s$:
\begin{enumerate}
	\item For each $i$ in $(1\text{, } \dots\text{, } n_x) $, sample $Z^x_i$ from a truncated normal distribution, where the lower threshold is $a^x_i$ given in \eqref{minthresh} and the upper threshold is $b^x_i$ given in \eqref{maxthresh}:
	
	\[ (Z_i^x  \mid z_{i'}^{x}\text{, } z_i^y\text{, }\rho_{z^x, z^y}) \sim \mathcal{N}_{\left(a_i^x\text{, } b_i^x\right)}\left(z_i^y \rho_{z^x\text{, } z^y}\text{, } \sqrt{1-\rho^2_{z^x, z^y}}\right) \]
	
	\item For each $i$ in $(1\text{, } \dots\text{, } n_y) $, the sampling procedure for $Z^y_i$ is analogous to step 1.
	\item Sample a new proposal for $\rho_{z^x\text{, } z^y}$, denoted $\rho^*$, from the asymptotic normal approximation to the sampling distribution of Fisher's $z$-transform of $\rho$ \citep{Fischer1915}:
	\[ \tanh^{-1}(\rho^*) \sim \mathcal{N}\left(\tanh^{-1}(\rho^{s-1}), \frac{1}{\sqrt{(n-3)}}\right) .\] 
	
	The acceptance rate $\alpha$ is determined by the likelihood ratio of $(z^x,z^y | \rho^*)$ and $(z^x\text{, }z^y \mid \rho^{s-1})$, where each likelihood is determined by the bivariate normal distribution in \eqref{bivTruncNorm}:
	\[ \alpha = \min\left(1\text{, }  \frac{P(z^x\text{, }z^y \mid \rho^*)}{P(z^x\text{, }z^y \mid \rho^{s-1})}  \right). \]
	
\end{enumerate}
Repeating the algorithm a sufficient number of times yields samples from the posterior distributions of  $z^x$, $z^y$, and $\rho_{z^x\text{, }z^y}$.

\subsection{Transforming Parameters}
The transition from Pearson's $\rho$ to Spearman's $\rho_s$ can be made using a statistical relation described in \cite{Kruskal1958}. This relation, defined as
\[  \rho_s = \frac{6}{\pi} \sin^{-1}\left(\frac{\rho}{2}\right). \]
enables the transformation of Pearson's $\rho$  to  Spearman's $\rho_s$ when the data follow a bivariate normal distribution. Since the latent data are assumed to be normally distributed, this means that the posterior samples for Pearson's $\rho$ can be easily transformed to posterior samples for Spearman's $\rho_s$.

\subsection{Simulation Study}

Similar to the previous tests, the behavior of the latent normal correlation test was assessed with a simulation study.
For four values of Spearman's $\rho_s$ (0, 0.3, 0.8) and three values of $n$ (10, 20, 50), $1{,}000$ data sets were generated under four copula models: Clayton, Gumbel, Frank, and Gaussian \citep{Sklar1959,Nelsen2006,Genest2007,Colonius2016}. Using Sklar's theorem, copula models decompose a joint distribution into univariate marginal distributions and a dependence structure (i.e., the copula). This decomposition enables the generation of data for specific values of Spearman's $\rho_s$. Furthermore, the copula is independent of the marginal distributions of the data and can therefore encompass a wide range of distributions.

\begin{figure}[h!]
	\centering
	\centerline{ 
		\includegraphics[trim=0cm 0cm 0cm 0cm,clip=true,scale=0.6]{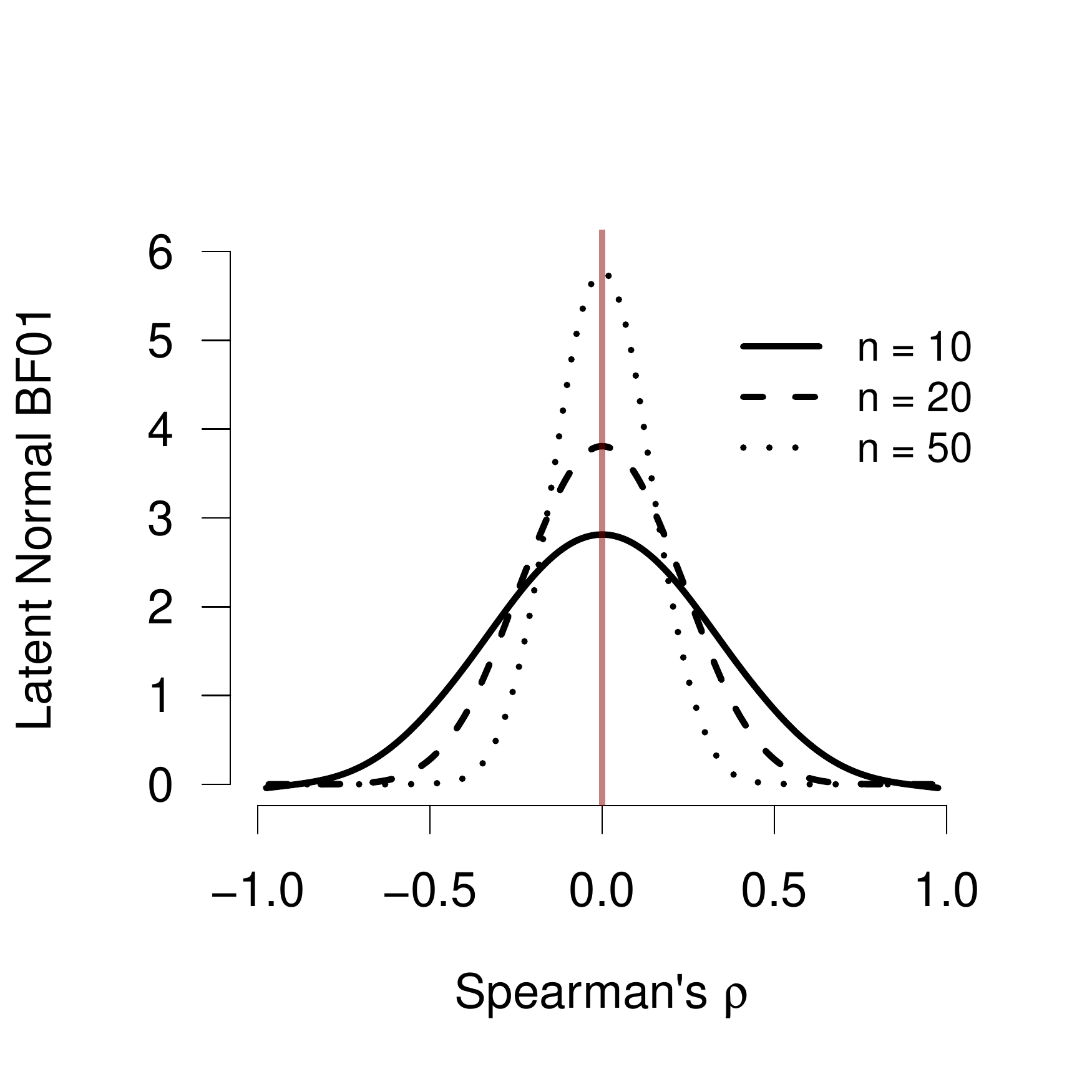}}
	\caption{The relationship between the latent normal Bayes factor and the observed rank-based test statistic is illustrated for data generated with the Clayton copula. The relationship is clearly defined, and maximum evidence in favor of $\mathcal{H}_0$ is attained when Spearman's $\rho_s = 0$. The further Spearman's $\rho_s$ deviates from 0, the stronger the evidence in favor of $\mathcal{H}_1$ becomes. The lines are smoothing splines fitted to the observed Bayes factors.}
	\label{fig:spearmanBF_Rho}
\end{figure}

Similar to the previous tests, the relationship between the latent normal Bayes factor and the observed rank-based statistic was analyzed. Figure \ref{fig:spearmanBF_Rho} illustrates this relationship, fitted with a cubic smoothing spline \citep{chambers1992}, for various values of $n$, for data generated with the Clayton copula. The relationship is similar to those shown for the previous tests: maximum evidence in favor of $\mathcal{H}_0$ is attained when the observed Spearman's $\rho_s$ equals $0$. The further the observed test statistic deviates from 0, the more evidence is accumulated in favor of $\mathcal{H}_1$. Furthermore, the obtained Bayes factor grows more decisive as $n$ increases. Highly similar results were obtained for the other copulas that were considered  (see the online supplementary material at \url{https://osf.io/gny35/} for the results of these scenarios).

\begin{figure}[h!]
	\centering
	\centerline{ 
		\includegraphics[trim=0cm 0cm 0cm 0cm,clip=true,scale=0.6]{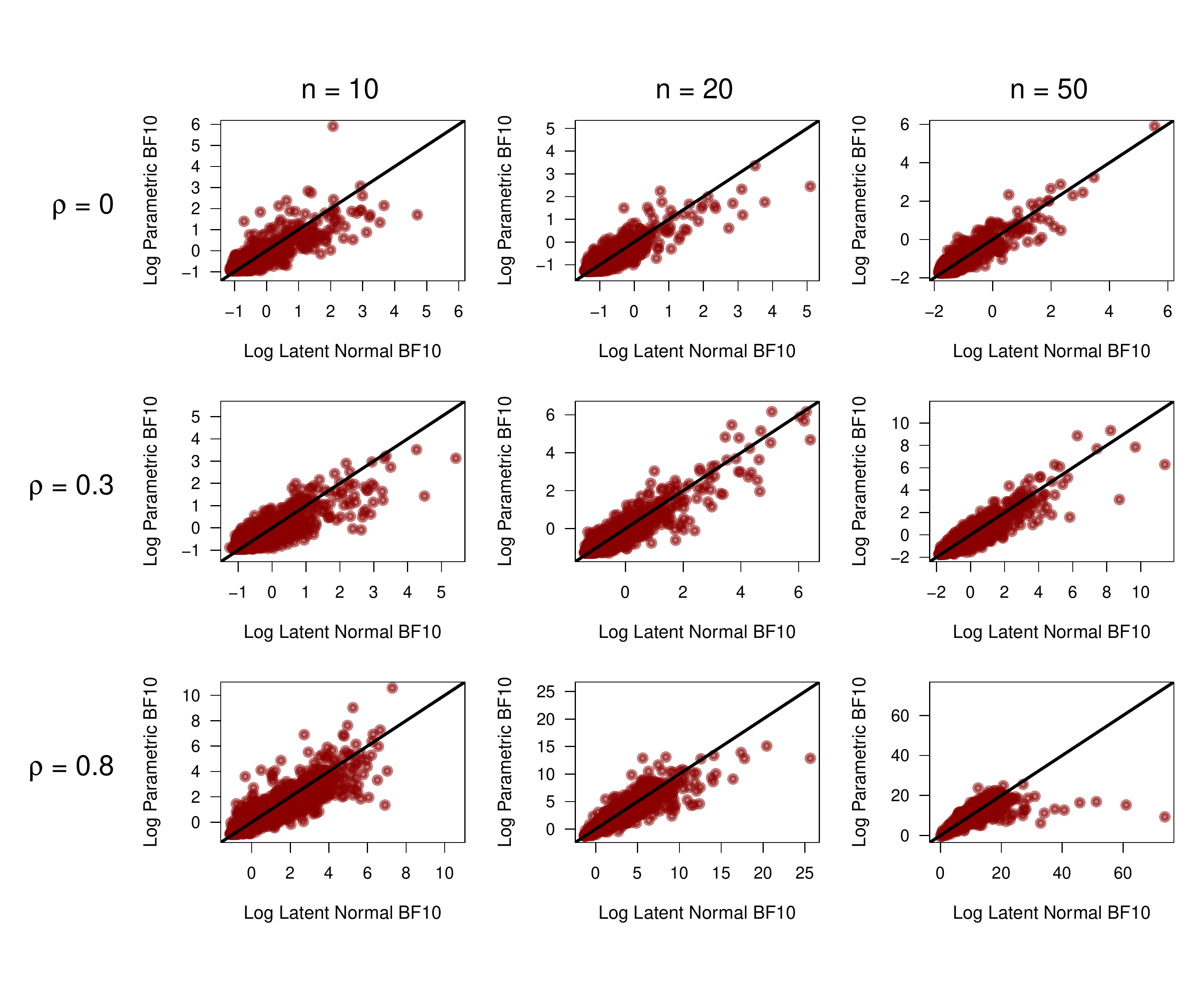}}
	\caption{For all combinations of Spearman's $\rho_s$ and $n$, the relationship between the latent normal Bayes factor and the parametric Bayes factor is shown for  data generated with the Clayton copula. The black lines indicate the point of equivalence. The two Bayes factors are generally in agreement.}
	\label{fig:spearmanBFcomp}
\end{figure}

Secondly, the relationship between the latent normal Bayes factor and the parametric Bayes factor \citep{lyetal2018}  for testing correlations was analyzed. Figure \ref{fig:spearmanBFcomp} shows the results for all combinations of $n$ and $\rho$ that were used, for the Clayton copula. The two Bayes factors are generally in agreement. An important remark here is that the marginal distributions of the data are not taken into account. The data generated with the copula method are located on the unit square, and if so desired, can then be transformed with the inverse cdf to follow any desired distribution. These transformations are monotonic, and therefore do not affect the rank-based Bayes factor, whereas the parametric Bayes factor can be heavily affected by this. This underscores an important property of the rank-based Bayes factor: it solely depends on the copula (i.e., the only component of the data that pertains to the dependence structure), and not on the marginal distribution of the data.

\subsection{Data Example}
We return to the data set from \citet{cortez2008} and examine the possibility that math grades (ranging from $0$ to $20$) are associated with the quality of family relations (self-reported on a Likert scale that ranges from $1-5$). The hypotheses are specified as follows, 
\[ \mathcal{H}_0 : \rho = 0, \]  
\[ \mathcal{H}_1 : \rho \sim \text{Uniform}[-1\text{, }1], \]
where the null hypothesis specifies the lack of an association between the two variables and the alternative hypothesis assigns the degree of association a uniform prior distribution (e.g., \citealp{Jeffreys1939}; \citealp*{lyetal2016}).

\begin{figure}[h!]
	\centering
	\centerline{ 
		\includegraphics[trim=0cm 0cm 0cm 0cm,clip=true,scale=0.6]{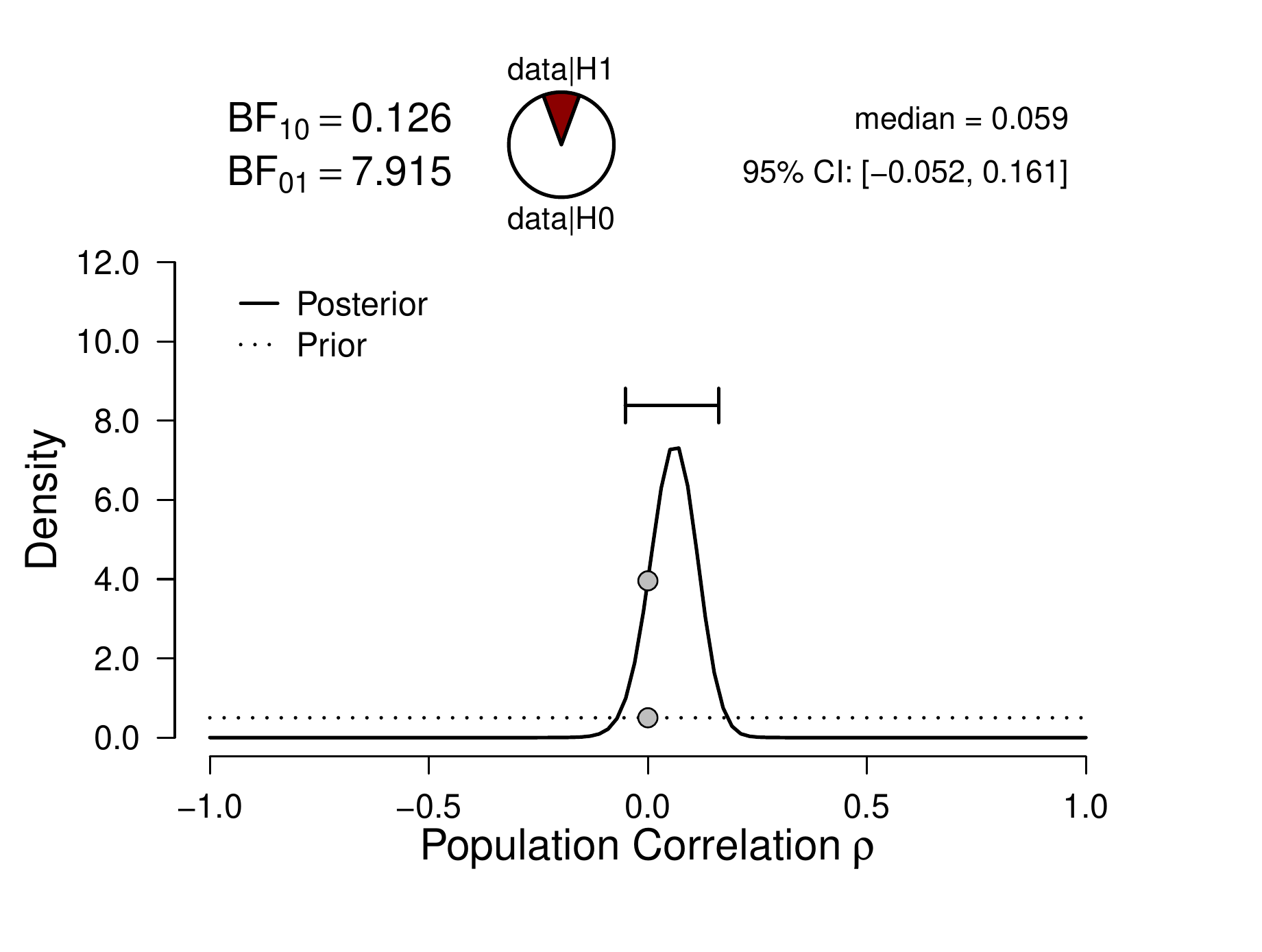}}
	\caption{Is performance on a math exam associated with the quality of family relations? Results for the Bayesian version of Spearman's $\rho_s$ as applied to the data set from \citet{cortez2008}. The dashed line indicates the uniform prior distribution, and the solid line indicates the posterior distribution. The two grey dots indicate the prior and posterior ordinate at the point under test, in this case $\rho = 0$. The ratio of the ordinates gives the Bayes factor.}
	\label{fig:rhoSampleData}
\end{figure}

Figure \ref{fig:rhoSampleData} shows the resulting posterior distribution for $\rho_s$ under $ \mathcal{H}_1 $ and the associated Bayes factor. The posterior median for $\rho_s$ equals $0.059$, with a $95\%$ credible interval that ranges from $-0.052$ to $0.161$. The corresponding Bayes factor indicates that the data are about $7.915$ times more likely under $ \mathcal{H}_0$ than under $\mathcal{H}_1$, indicating moderate evidence against an association between math performance and the quality of family ties.

\section{Concluding Comments}
This article outlined a general methodology for applying conventional Bayesian inference procedures to ordinal data problems. Latent normal distributions are assumed to generate impoverished rank-based observations, and inference is done on the model parameters that govern the latent normal level. This idea, first proposed by \citet{Pearson1900}, yields all the advantages of ordinal inference including robustness to outliers and invariance to monotonic transformations. Moreover, the methodology also handles ties in a natural fashion, which is important for coarse data such as provided by popular Likert scales. Furthermore, the robustness of the latent normal method is underscored by the simulation studies performed for each test. These results illustrate that the method provides accurate inference, even if the data are not normally distributed.

By postulating a latent normal level for the observed rank data, the advantages of ordinal inference can be combined with the advantages of Bayesian inference such as the ability to update uncertainty as the data accumulate, the ability to quantify evidence, and the ability to incorporate prior information. It should be stressed that, even though our examples used default prior distributions, the proposed methodology is entirely general in the sense that it also applies to informed or subjective prior distributions \citep{gronau2018informed}.

For computational convenience and ease of interpretation, our framework used latent normal distributions. This is not a principled limitation, however, and the methodology would work for other families of latent distributions as well (e.g., \citealp{Albert1992}). 

In sum, we have presented a general methodology to conduct Bayesian inference for ordinal problems, and illustrated its potential by developing Bayesian counterparts to three popular ordinal tests: the rank sum test, the signed rank test, and Spearman's $\rho_s$. Supplementary material, including simulation study results, R-code for each method and the example data used, is available at \url{https:https://osf.io/gny35/}. In the near future we intend to make these tests available in the open-source software package JASP (e.g., \citealp{JASP2018}; \url{jasp-stats.org}), which we hope will further increase the possibility that the tests are used to analyze ordinal data sets for which the traditional parametric approach is questionable.

\bibliographystyle{tfs}

\end{document}